\documentclass[%
 aip,
 amsmath,amssymb,
 reprint,%
]{revtex4-1}

\usepackage{graphicx}
\usepackage{dcolumn}
\usepackage{bm}

\usepackage[utf8]{inputenc}
\usepackage[T1]{fontenc}
\usepackage{mathptmx}
\usepackage{etoolbox}
\usepackage{xcolor}

\makeatletter
\def\@email#1#2{%
 \endgroup
 \patchcmd{\titleblock@produce}
  {\frontmatter@RRAPformat}
  {\frontmatter@RRAPformat{\produce@RRAP{*#1\href{mailto:#2}{#2}}}\frontmatter@RRAPformat}
  {}{}
}%
\makeatother
\begin{document}

\preprint{AIP/123-QED}

\title[Experimental validation of the diverse incident angle]{Experimental validation of the diverse incident angle performance of a pulse-width-dependent antenna based on a waveform-selective metasurface in a reverberation chamber}
\author{Kaito Ozawa}
\author{Hiroki Wakatsuchi}%
 \email{wakatsuchi.hiroki@nitech.ac.jp}
\affiliation{ 
Department of Engineering, Graduate School of Engineering, Nagoya Institute of Technology, Gokiso-cho, Showa, Nagoya, Aichi, 466-8555, Japan
}%

\date{\today}

\begin{abstract}
We present experimental validation results of a pulse-width-dependent antenna based on a waveform-selective metasurface that behaves differently according to the incoming waveform, more specifically, the incoming pulse width, even at a fixed frequency. This waveform-selective behavior is integrated with an antenna design to preferentially accept a predetermined incoming waveform while rejecting another waveform at the same frequency. In particular, we report how the waveform-selective antenna performance can be evaluated for diverse incident angles, which is enabled by a measurement method based on a reverberation chamber. This method facilitates variation of the electromagnetic fields and modes inside the chamber and mimicking of complicated wireless communication environments through rotation of the internal stirrer. The experimental results verify that our antenna design concept based on waveform-selective metasurfaces can filter different signals with a wide range of incident angles, opening the door to utilizing such pulse-width-dependent antennas to suppress electromagnetic noise in more realistic wireless communication environments. 
\end{abstract}

\maketitle

\begin{figure}[tb!]
\includegraphics[width=0.7\linewidth]{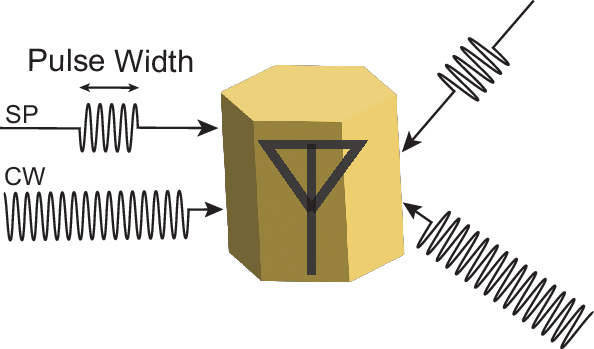}
\caption{\label{fig:1} Conceptual image of a pulse-width-dependent antenna based on a waveform-selective metasurface. The antenna is capable of distinguishing short pulses (SPs) from continuous waves (CWs) even if they enter from various angles. }
\end{figure}

Metasurfaces are two-dimensional structures that enable control of electromagnetic waves and related phenomena at will via artificially engineered subwavelength composite unit cells.\cite{MTMbookEngheta, EBGdevelopment, yu2011light, yu2014flat} For instance, metasurfaces readily provide abrupt phase changes for wavefront shaping and beamforming\cite{yu2011light, yu2014flat, pfeiffer2013metamaterial} and a high surface impedance for electromagnetic noise suppression and antenna performance enhancement.\cite{EBGdevelopment, CCemcBook} In particular, superior metasurface performance is attained by using nonlinearity,\cite{lapine2014colloquium} which breaks the constraints imposed by classic linear time-invariant (LTI) systems.\cite{antsaklis1997linear, takeshita2024frequency} For instance, metasurfaces designed with nonlinear circuit elements provide power-dependent beamforming\cite{luo2015self} and absorption capabilities\cite{li2017high, li2017nonlinear} and can be exploited as power limiters to prevent the propagation of high-power noises only while permitting the propagation of weak communication signals at the same frequency.\cite{aplNonlinearMetasurface, zhou2021high, wang2022design} More importantly, the time-harmonized response at a fixed frequency can be broken by using nonlinearity within metasurface unit cell design, which provides an additional degree of freedom to control electromagnetic waves and radio-frequency (RF) communication signals. A series of recent studies reported that metasurfaces, including diode bridges, successfully varied the time-domain response for signals at the same frequency in accordance with the incoming waveform, specifically the incoming pulse width.\cite{wakatsuchi2013waveform, wakatsuchi2019waveform, takeshita2024frequency} This waveform-selective response (or pulse-width-dependent response) has thus far been exploited to address a wide range of issues, including those in signal processing,\cite{f2020temporal} electromagnetic interference,\cite{wakatsuchi2019waveform}, sensing\cite{ushikoshi2023pulse}, and limited frequency resources.\cite{takeshita2024frequency} In particular, these waveform-selective metasurfaces play an important role in antenna design\cite{ushikoshi2018metasurfaces, ushikoshi2023pulse, vellucci2019waveform, barbuto2020waveguide, barbuto2021metasurfaces} to vary the radiation pattern and receiving performance without utilizing additional energy resources or frequency changes, thus expanding the design possibility of next-generation wireless communications. However, their performance has thus far only been evaluated at fixed incident angles, while the incident angle varies in realistic wireless communication environments. For this reason, this study reports the performance of a pulse-width-dependent antenna based on a waveform-selective metasurface for diverse incident angles (Fig.\ \ref{fig:1}). To facilitate the evaluation, we propose an experimental approach using a reverberation chamber, which contains a stirrer to readily change the internal electromagnetic field distribution and mode as well as the angle of the impinging wave. However, unlike ordinary free-space wave measurements, within the reverberation chamber, the received signal duration becomes much longer than the pulse width of the generated signal, which requires adjustment of the energy calculation and pulse period durations in the time domain. By properly considering these two factors, we experimentally verify that the pulse-width-dependent time-varying receiving performance can be obtained for a wide range of incident angles. 

\begin{figure}[tb!]
\includegraphics[width=\linewidth]{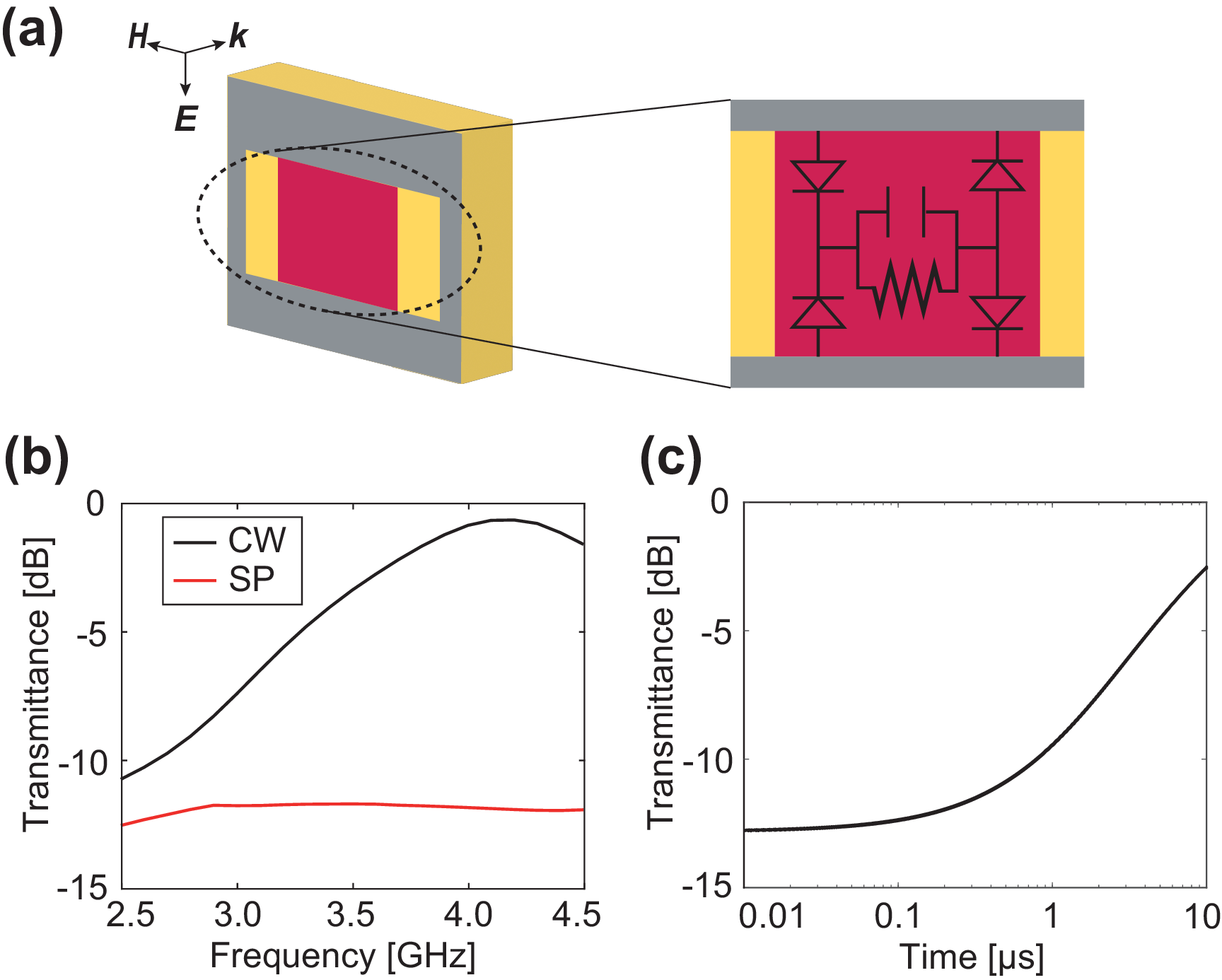}
\caption{\label{fig:2} Model of the periodic unit cell of the proposed waveform-selective metasurface and its simulation results. (a) The simulation model and loaded circuit elements. (b) Frequency-domain transmittances for 50-ns SPs and CWs. The input power was set to 10 dBm. (c) Time-domain transient transmittance. The frequency was fixed at 4.1 GHz. }
\end{figure}

In this study, the periodic unit cell design shown in Fig.\ \ref{fig:2}a was adopted.\cite{wakatsuchi2019waveform} Our unit cell was composed of a 1.27-mm-thick dielectric substrate (Rogers 3010) and a planar conducting layer with a rectangular slit (5 mm $\times$ 16 mm) and an 18-mm periodicity. Across the slit, a diode bridge (Avago, HSMS-286x series) was connected to fully rectify electric charges induced by incident waves, thus generating an infinite set of frequency components. However, the largest energy appeared at zero frequency, as expected from the Fourier expansion of $|\sin|$.\cite{wakatsuchi2019waveform} Therefore, as seen in the transients of classic DC circuits, the metasurface response became time-dependent by introducing both reactive and resistive circuit components and coupling the circuit response to the electromagnetic response of the metasurface. Specifically, we connected a capacitor (10 nF) to a resistor (100 k$\Omega$) within the diode bridge. In this case, the energy of a short pulse (SP) signal was fully stored in the capacitor. This energy was then strongly dissipated by the parallel resistor, so the transmission performance for SP signals was limited. However, the transmission performance was enhanced with an increase in the pulse width to a continuous wave (CW) even at the same frequency. In this case, the capacitor was fully charged such that the induced electric charges could not enter the diode bridge. Therefore, the intrinsic resonant mechanism of the metasurface unit cell (or slit structure) was maintained to strongly transmit CWs. 

This waveform-selective metasurface transmitting long pulses (or CWs) was used as part of the antenna design. First, a monopole receiver (20 mm tall) was set at the center of a ground plane and then surrounded by six metasurface panels, each of which contained 2 $\times$ 3 unit cells, to configure a hexagonal prism structure\cite{ushikoshi2023pulse} (the closest distance between the inner surface of the metasurface panels and the center of the grounded monopole was approximately 30 mm). Without the metasurface panels, the monopole antenna was expected to show a time-invariant omnidirectional radiation pattern or receiving profile, which could be changed by the time-varying electromagnetic response of our metasurface. 

We evaluated the receiving performance of the metasurface-based antenna for various incident angles. In conventional methods, the angle of incident waves needs to be changed in a three-dimensional space. However, such approaches require not only an extremely long time to mechanically vary the physical location of the transmitting antenna (or rotate the receiver) but also a large space to ensure a far-field distance. Hence, we adopted a different approach based on a reverberation chamber (Toyo EMC Engineering, Model-5955). This method facilitated mimicking of complicated wireless communication environments through rotation of a conducting stirrer that largely changed the incident wave angle and the internal electromagnetic fields in accordance with the stirrer rotation angle. In this approach, the metasurface-based antenna was used as a receiver, and a standard horn antenna (Corry Micronics Inc., CMILB-284-10-C-S) was used as a transmitter whose signals were scattered by the above stirrer and eventually received by the metasurface-based antenna. The transmittance (or the received power) was calculated by dividing the transmitted (received) energies of all the stirrer rotation angles (0 to 359 degrees in 1-degree steps) by the incident energies. 

Before we evaluated the diverse incident angle performance of our metasurface-based antenna, we simulated the transmittance of the metasurface unit cell with periodic boundaries, as shown in Fig.\ \ref{fig:2}a. This model was simulated using a co-simulation method available in ANSYS (version 2022).\cite{wakatsuchi2019waveform} The transmittances for 50-ns SPs and CWs when the input power was set to 10 dBm were obtained, as shown in Fig.\ \ref{fig:2}b. According to these results, the transmittance for CWs was more enhanced than that for SPs near 4.1 GHz, which can be explained by the abovementioned waveform-selective mechanism. To more precisely evaluate this mechanism, we plotted the transmittance at 4.1 GHz in the time domain, as shown in Fig.\ \ref{fig:2}c. This figure clearly demonstrates that the transmittance gradually increased between 100 ns and 10 $\mu$s from -12.4 to -2.5. This time-varying electromagnetic response can readily be adjusted by changing the time constant and the circuit configuration.\cite{aplEqCircuit4WSM, wakatsuchi2015time} 

\begin{figure*}[tb!]
\includegraphics[width=\linewidth]{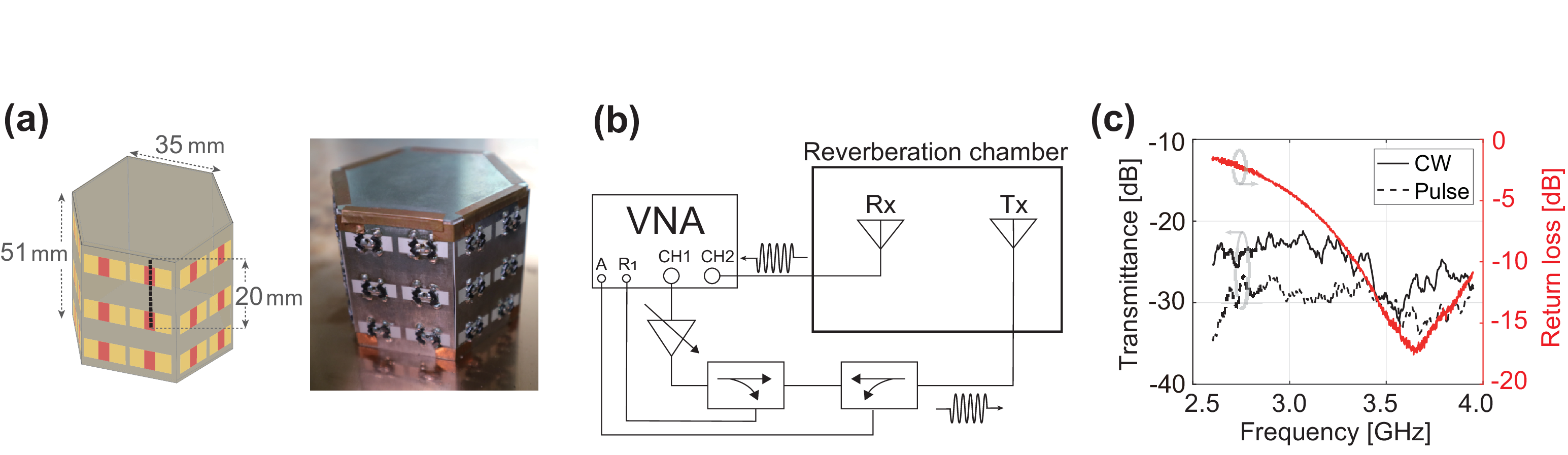}
\caption{\label{fig:3} Pulse-width-dependent metasurface-based antenna and VNA measurement results. (a) Antenna design. The grounded monopole represented by the dashed black line is covered by six panels of the waveform-selective metasurface and a hexagonal conducting plate. (b) Measurement system. (c) Results including the return loss of the bare monopole receiver.}
\end{figure*}

Next, as shown in Fig.\ \ref{fig:3}a, this waveform-selective metasurface was used to assemble the pulse-width-dependent antenna described above. This antenna was deployed in the abovementioned reverberation chamber connected to a vector network analyzer (VNA) (Keysight Technologies, N5249B) to experimentally evaluate the transmission performance, as shown in Fig.\ \ref{fig:3}b. Here, the input signal was increased to 35 dBm by an amplifier (Ophir, 5193RF) to ensure that the input power was large enough to turn on the diodes loaded on the antenna. The measurement results are shown in Fig.\ \ref{fig:3}c. This figure shows that the transmittance for SPs was entirely lower than that for CWs. Importantly, however, the bandwidth was much wider than that shown in the numerical simulation results of Fig.\ \ref{fig:2} as well as the theoretical maximum bandwidth of ordinary metasurface absorbers.\cite{rozanov2000ultimate} Also, note that this bandwidth was wider than that of the return loss of the bare monopole receiver (see the red curve in Fig.\ \ref{fig:3}c). The reason for this wider bandwidth is further analyzed in the following part of this paper. 

\begin{figure*}[tb!]
\includegraphics[width=\linewidth]{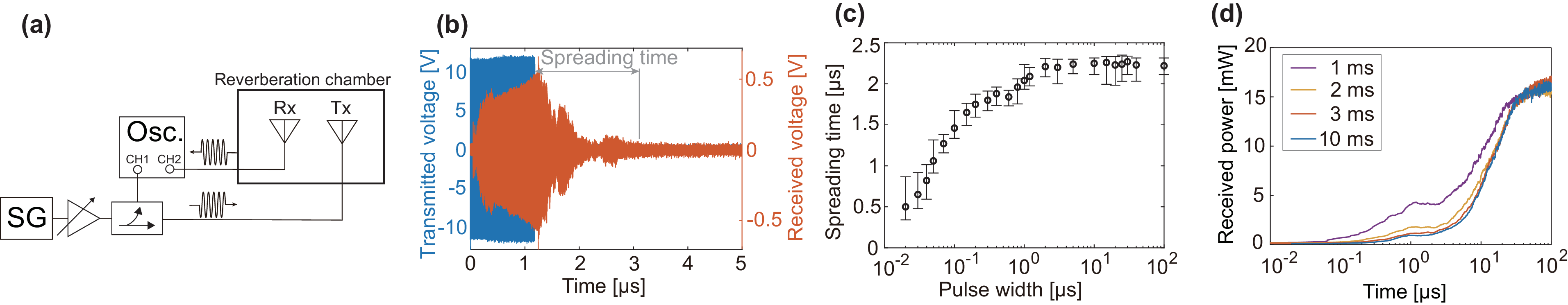}
\caption{\label{fig:4} Time-domain measurement results. (a) Measurement system. (b) Extended waveform tail in the reverberation chamber. (c) Spreading time as a function of the pulse width. Each symbol represents the average result, including all of the rotation angles of the stirrer. Their variations are shown by the vertical bars. (d) Transient transmittance in the time domain for various pulse periods. }
\end{figure*}

To more precisely evaluate the performance of the antenna in the time domain, we changed part of the measurement setup, replacing the VNA with a signal generator (Anritsu, MG3690C) and an oscilloscope (Keysight Technologies, DSOX6002A), as shown in Fig.\ \ref{fig:4}a. First, by using a 35-dBm input power at 3.00 GHz, we generated a 1.2-$\mu$s pulse and obtained the incident and transmitted waveforms shown in Fig.\ \ref{fig:4}b. This figure indicates that while the incident pulse width was set to 1.2 $\mu$s, the waveform received by the metasurface-based antenna (i.e., the transmitted waveform) continued for much longer and included a nonnegligible waveform tail, which was presumably because of a large number of scattering events within the reverberation chamber. Thus, we evaluated how this spreading time denoted by the gray arrow of Fig.\ \ref{fig:4}b was influenced by the incident pulse width. The results are plotted in Fig.\ \ref{fig:4}c as a function of the incident pulse width. In this case, we set the pulse period to a sufficiently large value, specifically 50 ms. According to Fig.\ \ref{fig:4}c, the spreading time was shorter than 2.4 $\mu$s, which indicated that the transmitted energy must be evaluated over a sufficiently long period of time. Note that the long waveform tail could not fully be captured in the first measurement method using our VNA. In addition to the pulse spreading issue, we also investigated the proper pulse period, as shown in Fig.\ \ref{fig:4}d, in which the input pulse width was set to 100 $\mu$s for various pulse periods. Note that this measurement was important as the waveform-selective metasurface did not instantaneously discharge the stored energy. As a result, Fig.\ \ref{fig:4}d showed that when a pulse period longer than 3 ms was used, the transient transmittance did not vary with the pulse period. Based on these results, in the following measurements, 3 ms was adopted as a sufficiently long pulse period. 

\begin{figure}[tb!]
\includegraphics[width=\linewidth]{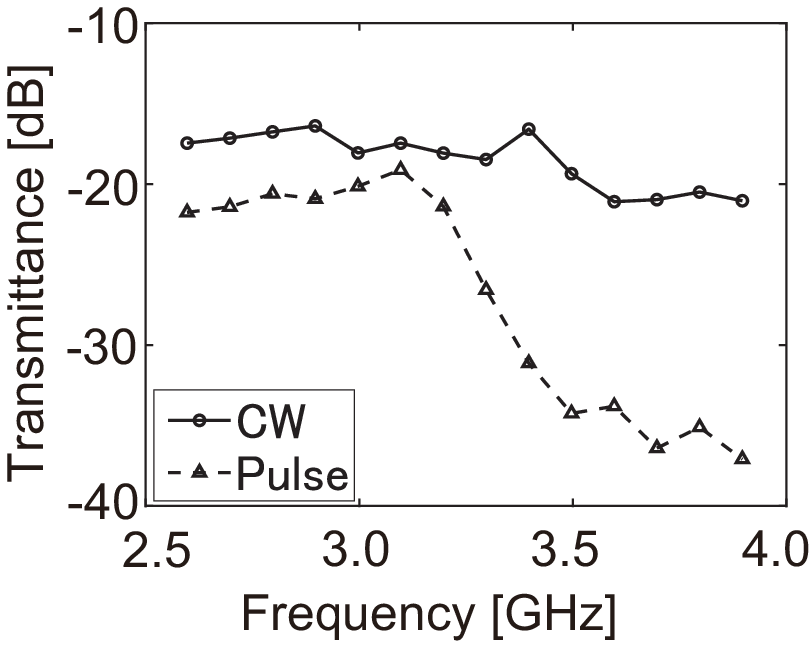}
\caption{\label{fig:5} Transmittances obtained using the measurement method shown in Fig.\ \ref{fig:4}.  }
\end{figure}

Considering the measurement results in Fig.\ \ref{fig:4}, we conducted additional time-domain measurements while changing the incident frequency to properly evaluate the pulse-width-dependent performance of our metasurface-based antenna. The results are shown in Fig.\ \ref{fig:5}, and the transmittance for CWs was approximately 10 dB higher than that for 50-ns SPs only around 3.7 GHz. Note that these results completely differed from those in Fig.\ \ref{fig:3}c, where the entire transmitted waveform (or energy) was not fully captured. Additionally, the bandwidth of our metasurface-based antenna in Fig.\ \ref{fig:5} was no longer unreasonably wide. Moreover, this figure supports that our metasurface-based antenna works at diverse incident angles, with the receiving performance still being varied in accordance with the incident pulse width, which can be utilized as an additional degree of freedom to control electromagnetic waves and related phenomena even at a fixed frequency. 

In conclusion, we demonstrated a pulse-width-dependent antenna based on a waveform-selective metasurface that varies the electromagnetic response for signals even at the same frequency and more strongly transmits CWs than SPs. In particular, we proposed a measurement method using a reverberation chamber that allowed complicated electromagnetic field distributions with various incident angles to be readily mimicked. Importantly, however, this method turned out to extend the pulse waveform tail and to require a sufficiently long pulse period, which could not be satisfied with our VNA-based measurement approach. We therefore conducted additional time-domain measurements using a signal generator and oscilloscope pair. Under these conditions, we successfully determined the experimental receiving performance of our metasurface-based antenna. Our results support that the antenna design based on waveform-selective metasurfaces can preferentially accept signals from a wide range of incident angles. Hence, our study opens the door to utilizing antennas based on waveform-selective metasurfaces in more realistic wireless communication environments, together with the pulse-width dependence as an additional degree of freedom to control electromagnetic waves and fields even at a constant frequency despite rigorously regulated frequency assignments. 

\begin{acknowledgments}
This work was supported in part by the Japan Science and Technology Agency (JST) under Precursory Research for Embryonic Science and Technology (PRESTO) No.\ JPMJPR193A and under Fusion Oriented Research for Disruptive Science and Technology (FOREST) No.\ JPMJFR222T and by the National Institute of Information and Communications Technology (NICT), Japan under commissioned research No.\ 06201.
\end{acknowledgments}

\section*{Data Availability Statement}

The data that support the findings of this study are available from the corresponding authors upon reasonable request.

\providecommand{\noopsort}[1]{}\providecommand{\singleletter}[1]{#1}%

\end{document}